%% file: arxiv-version.tex
\documentclass[11pt]{llncs}
\usepackage{geometry}
\usepackage{Xiao-llncs}

\usepackage{xcolor}
\usepackage{tabularx}
\usepackage{booktabs}
\usepackage{numprint}
\usepackage[notextcomp,notext]{stix}

\input{macros}

\usepackage[absolute]{textpos}
\makeatletter
\def\@fnsymbol#1{\ensuremath{\ifcase#1\or *\or \dagger\or \ddagger\or
   \mathsection\or \mathparagraph\or \|\or **\or \dagger\dagger
   \or \ddagger\ddagger \else\@ctrerr\fi}}
\makeatother

\begin{document}
\begin{textblock}{100,90}(14,3)
\noindent A preliminary version of this paper appears in the proceedings of the \textit{22nd Privacy Enhancing Technologies Symposium} (\textsc{PETS 2022}). 
\end{textblock}

\title{~\\\LARGE{SoK: Plausibly Deniable Storage}}

\author{
Chen Chen\inst{1}
\and 
 Xiao Liang\inst{1}
\and 
Bogdan Carbunar\inst{2}
\and 
Radu Sion\inst{1}
}

\institute{
 Stony Brook University, NY, USA\\ 
 \email{\{cynthia.us12,xiao.crpyto\}@gmail.com},~~
 \email{r@zxr.io}\\[0.5em]
 \and
 Florida International University, FL, USA\\
 \email{carbunar@gmail.com}
}

\let\oldaddcontentsline\addcontentsline
\def\addcontentsline#1#2#3{}
\maketitle
\def\addcontentsline#1#2#3{\oldaddcontentsline{#1}{#2}{#3}}

\begin{abstract}
Data privacy is critical in instilling trust and empowering the societal pacts of modern technology-driven democracies. Unfortunately, it is under continuous attack by overreaching or outright oppressive governments, including some of the world's oldest democracies. Increasingly-intrusive anti-encryption laws severely limit the ability of standard encryption to protect privacy. New defense mechanisms are needed.\\[-0.5em]

{\em Plausible deniability} (PD) is a powerful property, enabling users to hide the existence of sensitive information in a system under direct inspection by adversaries. Popular encrypted storage systems such as TrueCrypt and other research efforts have attempted to also provide plausible deniability. Unfortunately, these efforts have often operated under less well-defined assumptions and adversarial models. Careful analyses often uncover not only high overheads but also outright security compromise. Further, our understanding of adversaries, the underlying storage technologies, as well as the available plausible deniable solutions have evolved dramatically in the past two decades. The main goal of this work is to systematize this knowledge. It aims to: 
\begin{enumerate}
\item identify key PD properties, requirements, and approaches; 
\item present a direly-needed unified framework for evaluating security and performance;
\item explore the challenges arising from the critical interplay between PD and modern system layered stacks; 
\item propose a new ``trace-oriented'' PD paradigm, able to decouple security guarantees from the underlying systems and thus ensure a higher level of flexibility and security independent of the technology stack.
\end{enumerate}
\vspace{0.5em}

This work is meant also as a trusted guide for system and security practitioners around the major {\em challenges} in understanding, designing, and implementing plausible deniability into new or existing systems.
\end{abstract}

\setcounter{tocdepth}{3}
\tableofcontents
\clearpage

	\pagenumbering{arabic}

\input{introduction}


\section{Model}
\label{section:single-vs-multi}


In this section, we provide the problem setup for PD, then describe the system and adversary model.


\subsection{The Plausible Deniability Problem}

Plausibly deniable storage systems need to allow users to store public data, {\em and} sensitive hidden data. Public data does not require protection, and is potentially known by the adversary. Hidden data needs to be protected against coercive adversaries who can compel the user to hand over secret information (e.g.\ encryption keys). Under duress, the user may need to provide some information (e.g.\ keys to public data) that dismisses the adversary's suspicions, while most importantly denying the existence of the hidden data.



\subsection{System Model}
\label{section:model:system}

Modern storage systems comprise multiple layers that link the physical storage medium and the user applications. Example layers include the file system (FS), block device layer (BD), device mapper, flash translation layer (FTL), and the physical device, e.g., NAND flash or block device. The File System (\FS) layer is mandatory. It organizes data as files for better management. The Block Device (\BD) layer provides abstractions for block devices and maps multiple ``virtual volumes'' onto one block device, where a volume can be, for instance, a file system. It is optional and needed only if a block device is deployed as the storage media (e.g., the device mapper in the Linux kernel). Another optional layer is the Flash Translation (\FT) layer, needed if NAND flash is used as the physical device.

Not every storage system contains all the above layers. For example, if the physical storage medium is a NAND flash, then the storage system could consist of an \FS layer only, or both an \FS layer and an \FTL layer.

\para{Operation Traces}.
The storage device allows \Read and \Write operations. An operation trace is an ordered sequence of operations the system performs on the physical device, independent of layers or device properties. For instance, the operation trace (\Read, $l_1$), (\Write, $l_1$, $d_1$), (\Write, $l_2$, $d_2$) first reads the data from location $l_1$, then writes data $d_1$ into $l_1$ and $d_2$ into location $l_2$.

In the following, we assume that \Read operations do not modify the storage medium. Since only \Write operations leave traces on the storage medium, \Write is the only type of operation that the trace-oriented definition introduced in \Cref{sec:unified-def-mainbody} needs to consider.

\para{Mount and Unmount}.
Several PD systems require users to \Mount and \Unmount volumes or partitions in order to switch between accessing public and hidden data~\cite{jia2017deftl,chen2020infuse,chen2020pearl}, and even require them to \Unmount the hidden partition before handing the device to the adversary.

\subsection{Adversary Model}
\label{section:model:adversary}

We first detail common adversary assumptions. We then provide a classification of adversaries based on their capabilities, and introduce a novel, trace-oriented adversary. Further, we describe the standard, CPA game-inspired, plausible deniability definition.

\para{Adversary Assumptions and Capabilities}.
Adversaries are assumed to be able to access the device of a user, and attempt to compromise deniability, i.e., determine if the user is storing any secret data. Adversaries are generally assumed to be computationally bounded.

Adversaries are assumed to know the design of the deployed PD solution. They are also assumed not to know how many keys are used in the system, and to not have access to hidden user passwords or encryption keys. However, they can request the user to reveal passwords and encryption keys. The user is assumed in this case to reveal public passwords (including providing root privilege) and public keys. Adversaries can use such information to access and decrypt stored data. In addition, adversaries can use password cracking programs and perform forensics on the disk image.

Additional assumptions have been introduced to accommodate the applications or to trade for better performance. Those that significantly affect the design choices for PD are discussed as follows.
\begin{itemize}

\item
{\bf A1:} Adversaries are rational. Namely, an adversary will stop further coercion if it cannot prove the existence of any unrevealed data.

\item
{\bf A2:} Adversaries cannot observe run-time system state (e.g., DRAM, caches).

\item
{\bf A3:} Adversaries cannot perform malicious code injection on the system used by the user.

\end{itemize}

\noindent
Assumption A1 draws a line on the adversary's coercive behavior, and was made (sometimes implicitly) in the majority of exiting work. Assumptions A2 and A3 limit PD to disk states only. This captures a wide class of application scenarios, including the motivational examples in the introduction.

\para{Adversary Classification}.
Adversaries can be classified based on the data they can access on the user device:
\begin{itemize}

\item
{\bf Snapshot-Oriented Adversary}.
The typical adversary is snapshot-oriented. Such an adversary can only access snapshots of the physical device.

\item
{\bf Trace-Oriented Adversary}.
We introduce a novel, trace-oriented adversary, that can access not only device snapshots, but also the operation traces (\Cref{section:model:system}) that produce them.

\end{itemize}

\noindent
Traces are the result of probabilistic polynomial time (PPT) run-time computations on user requests, i.e., sequences of compliant logical instructions to be executed at a layer (\Cref{section:model:system}). For instance, traces at the BD layer traces may include block \Read and block \Write instructions, while at the FTL layer, traces may include page \Read, page \Write, and block $\Erase$ instructions. This is in contrast to run-time system state that includes the contents of memory and caches.

An example where an adversary can capture trace data is in flash. SSDs implement an \FTL layer inside-the-box that {\it sees} all operation traces (e.g., which inode pages are updated) before they are executed on the actual flash cells. However, the complex wear-leveling logic inside the \FTL maintains state both as meta information and on the device itself (e.g., un-mapped not-yet-{$\Erase$}d blocks containing compromising old data) that, when inspected, can directly reveal critical information about past traces or even the traces themselves.

Snapshot-oriented adversaries can be further classified based on their number of opportunities to inspect the user device:
\begin{itemize}

\item
{\bf Single-Snapshot Adversary}.
Such an adversary can see the device only once before eventually confronting the user and demanding access to information. This makes the design of efficient PD schemes significantly easier. Indeed, a single snapshot (i.e., of a randomized encrypted device) does not leak much (if any) information beyond its size. For PD then, it may be sufficient to hide the sensitive data ``encrypted''\footnote{The double quotation marks are due to the fact that although most schemes use standard encryption, there are some schemes (e.g.~\cite{anderson1998steganographic}) using primitives such as secret-sharing instead.}, e.g., indistinguishable from random ``free'' device areas.

\item
{\bf Multi-Snapshot Adversary}.
Such an adversary can take multiple snapshots of the device at different time points \cite{Czeskis:2008:DED:1496671.1496678}. Examples multi-snapshot adversaries include customs officers or hotel personnel with regular access. Data center servers may also face inspections by overreaching authorities empowered by rubberhose or ill-devised laws. For multi-snapshot PD, it is exponentially more difficult to balance the security-efficiency trade-off. Exploring this will be one of the main themes of \Cref{section:depedence-on-layers,section:comparison}.

\end{itemize}

\para{Standard, CPA-Game for PD}.
We now briefly describe the first formal definition of PD introduced by Blass et al.~\cite{blass2014toward} and refined in \cite{chakraborti2017datalair}. The definition of PD is provided through a cryptographic game, analogous to the one used to define encryption against {\em chosen-plaintext attacks} (CPA). We expand this to provide a unified definition of PD in \Cref{sec:unified-def-mainbody}.

The security game is played between a coercive adversary $\Adv$ and a challenger $\Challenger$ running the underlying PD scheme $\Sigma$. The adversary holds the credentials needed to access the public data, but is ignorant of the ones for hidden data. At the beginning, $\Challenger$ picks a random bit $b \pick \bits$. $\Adv$ is allowed to interact with $\Challenger$ for polynomial-many rounds. In each round, $\Adv$ issues access patterns $\Pat^0$ and $\Pat^1$ that share the same access requests to public data, but may contain different access requests to the hidden data. $\Challenger$ will always execute $\Pat^b$. At the end of these interactions, $\Adv$ gets the snapshot of the physical device. $\Adv$ wins if it can guess the value $b$ correctly. The scheme $\Sigma$ is said to achieve single-snapshot PD if the winning probability of $\Adv$ is $\le \frac{1}{2} + \negl(\SecPar)$, where $\negl(\SecPar)$ is a {\em negligible function} on the security parameter $\SecPar$.\footnote{The term $1/2$ reflects the fact that the $\Adv$ can guess randomly and win the game with probability $1/2$.} This game extends to capture multi-snapshot security by allowing $\Adv$ to access the device state at the end of each round.

\section{Unified PD Definition}
\label{sec:unified-def-mainbody}

\subsection{Independence of Storage Layers}
\label{section:depedence-on-layers}

The layered structure of modern storage systems (\Cref{section:model:system}) complicates the design of PD schemes. Yet, this is often overlooked and has not been studied in a systematic way. In the following we investigate how the security of PD solutions is affected by storage layers. We also introduce a new ``trace-oriented'' definition for PD. In the standard PD definition (\Cref{section:model:adversary}) the adversary gets to see snapshots of devices; A trace-oriented notion allows the adversary to also learn operation traces. Trace-oriented PD provides stronger security guarantees and more flexible deployment choices due to its reduced dependence on storage layers.

\para{Layer-Specific PD Solutions are Vulnerable}.
Because of the layered nature of modern technology stacks, PD solutions are often designed for a target layer $\Lay$, e.g., $\Lay$ could be $\FS$, $\BD$ or $\FTL$. Then, in the security analysis, even the very existence of the underlying layers is often simply ignored. Unfortunately this results in designs that can be easily compromised by an adversary with access to operation traces (see example of trace-oriented compromise in \Cref{section:model:adversary}).

In general, an adversary with visibility into the state of one or more other layers, can compromise single-layer designs since that state can reveal access patterns and other security-sensitive information that a single-layer model simply cannot consider.

\para{Trace-Oriented PD: Removing Layer Dependency}.
\label{sec:trace-oriented_PD}
The above discussion leads to the following question: 
\begin{quote}
{\em Is it possible to achieve a stronger PD whose security is independent of individual technology stack layers?}
\end{quote}

Layer-independence is preferable. First, it enables modularity and ensures across-the-layers security. Second, it enables the evaluation of different schemes based on overall security strength. Performance metrics (e.g. time/space efficiency) also make better sense when they are least interwoven with stack layers. Otherwise, it is difficult to compare PD solutions operating on two different layers. Third, the fewer dependencies on implementation specifics, the better the security abstraction. PD can now be compared with other security constructs such as ORAMs; Such a connection is hard to establish for layer-dependent PD.  


We can then define trace-oriented plausible deniability by modifying the CPA-style security game of \Cref{section:model:adversary} in the following way: instead of device snapshots, the adversary will receive the operation traces as the reply to its challenge requests in the security game (along with the device snapshots). Namely, it is stipulated that the adversary cannot tell which of the two challenge sequences were executed, even if it gets to learn the outputs of the PD logic (aka operation traces) before they are physically executed on the storage medium. Intuitively, this is a stronger requirement than that of standard PD because operation traces may contain more information than snapshots---it is totally possible that two different sequences of operation traces lead to the same snapshot.

Operation traces are comprised of \texttt{Read} and \texttt{Write} operations. As mentioned in \Cref{section:model:system}, only \texttt{Write} operations leave traces on the storage medium. Thus, \texttt{Write} is the only type of operation that the trace-oriented definition needs to consider. Namely, it only requires that the \texttt{Write} traces reveal no information of the access requests to a PD scheme. Removing \texttt{Read} operations from traces is also preferable because an analogue of \Cref{lemma:tPD_wORAM} will show that including $\mathtt{Read}$ will lead to a trace-oriented PD definition that is equivalent to ORAMs (instead of write-only ORAMs), thus suffering ORAMs' efficiency lower-bounds \cite{goldreich1996software,boyle2016there,larsen2018yes,weiss2018there,hubavcek2019stronger}.

To achieve this we consider a function WOnly($\cdot$) that filters out the \texttt{Read} operations but passes the \texttt{Write} operations; The above security game can then be modified to return to the adversary the result of applying WOnly($\cdot$) on operation traces. This constitutes the final definition of trace-oriented PDs.

A PD scheme meeting the trace-oriented definition also satisfies the standard, CPA-style PD definition of \Cref{section:model:adversary}. Indeed, \Write traces (the output of $\mathsf{WOnly}$) contain all the information to induce storage medium snapshots; If they are oblivious of the input access request, so are the snapshots. Furthermore, it resolves the issue of layer dependency: notice that lower-layer traces are always obtained from higher-layer traces, via an implementation-specific PPT procedure. Since indistinguishable operation traces remain indistinguishable after being processed by arbitrary PPT procedures, trace-oriented PD schemes allow the existence of extra layers between the PD logic and the physical devices.
%

\para{Equivalence between Trace-Oriented PDs and Write-Only ORAMs.} 
Blass et al. \cite{blass2014toward} constructed a trace-oriented PD scheme from $\wORAM$. Further, in \Cref{sec:wORAM_from_PDs} we show that $\wORAM$ can also be constructed from trace-oriented PDs. This implies the following lemma:

\begin{lemma}\label{lemma:eq:woram-tpd}
Write-only ORAMs are both {\em sufficient and necessary} for trace-oriented PDs.
\end{lemma}

\subsection{Unified Definition}
\label{section:definition}

The CPA-Game for PD from \cite{blass2014toward,chakraborti2017datalair} defined in \Cref{section:model:adversary} is deeply integrated with the underlying application. New solutions have to repurpose this game to define PD at different system layer with specific underlying devices. Further, several constructions restricted the adversary's power in exchange for better efficiency, making it unclear how they fit into this game definition.

In this section we introduce a unified definition that 
\begin{enumerate}
	\item
	generalizes the CPA game in \Cref{section:model:adversary}, thus inherits all its advantages, e.g.~secure against CPA-style coercion attacks, applicable for both multi-snapshot and single-snapshot settings; 
	\item
	it encompasses existing constructions and admits comparisons among them (shown in \Cref{section:comparison}); 
	\item it can be instantiated for both the traditional device-oriented security model and the trace-oriented one proposed in \Cref{section:depedence-on-layers}.
\end{enumerate} 

We present the definition for both device-oriented and trace-oriented settings, with multi-snapshot adversaries. We use the parameter $\Lay$ (e.g., $\Lay$ can be $\FS$, $\BD$, $\FTL$) to restrict the game to the scenario where the adversary is attacking a storage device used at layer $\Lay$ (i.e., the device is directly connected to the layer $\Lay$).
%

The security game captures restrictions on the adversary's power through two parameters: (1) the number of rounds $r$ (single-snapshot when $r = 1$, multi-snapshot when $r> 1$), and (2) a new parameter $\Rule$, which can be instantiated by the designer. This leads to a more unified point of view, as all PD schemes indeed share the same abstraction modulo the parameters $\Lay$ and $\Rule$. Further, comparisons of security strength among different schemes become possible by investigating the restrictiveness of their respective $\Rule$ parameters. 

\para{Layer-Specific Notations.}
%
%
An $\Lay$-request is a legitimate access ($\Read$ or $\Write$) request to layer $\Lay$. An $\Lay$-pattern $\Pat$ is an ordered sequence of $\Lay$-requests. Let $\Pat_1 \cup \Pat_2$ denote the concatenation of requests in patterns $\Pat_1$ and $\Pat_2$.
%
%
We define the function OpTrace($\Lay, \Pat$) that, for a layer $\Lay$ and access request $\Pat$, outputs a sequence of operations that are meant to be executed on the underlying physical device (i.e., the ``{\em operation traces}'').

\begin{definition}
For a device $\Dev$ and layer $\Lay$, an {\em $\Lay$-layer PD} ($\LPD$) scheme $\Sigma$ consists of the following two algorithms $(\Setup, \Oper)$:
\begin{itemize}
    \item {\bf $\Setup(\SecPar, \Dev)$}:
    this function provides the initial setups. It takes as input the security parameter $\SecPar$ and the device $\Dev$. It outputs the tuple $(\Dev_\init, \Key_\pub, \Key_\hid)$, where $\Dev_\init$ is the initialized device, $\Key_\pub$ is the key used to protect the public data and $\Key_\hid$ is the key used to protect the hidden data.    
    \item {\bf $\Oper(\Dev_\state, \Pat, \Key_\pub, \Key_\hid)$}:
    $\Oper$ is a {\em stateful algorithm}, i.e., it may maintain {\em internal} state across consecutive invocations\footnote{This internal state should not be confused with the device state $\Dev_\state$ in the input to $\Oper$.}. It takes as input the current state $\Dev_\state$, an \Lay-pattern $\Pat$, and the key-pair $(\Key_\pub, \Key_\hid)$. If $\Pat$ is not a valid $\Lay$-pattern, the algorithm outputs $\bot$ and halts; Otherwise, it generates a new state $\Dev'$ accordingly, and updates the current state to $\Dev_\state \coloneqq \Dev'$. It outputs the updated state $\Dev_\state$:
    $$\Dev_\state \leftarrow \Oper(\Dev_\state, \Pat, \Key_\pub, \Key_\hid).$$
\end{itemize} 
\label{def:dpd}
\end{definition}


\begin{figure}[!tb]
    \begin{center}
    \fbox{
    \begin{minipage}{\columnwidth-10pt}
    Denote this game as $\PD^{\Lay}_{\Sigma,\Adv}(\SecPar, r)$. It is parameterized by a security parameter $\SecPar$, an \LPD scheme $\Sigma=(\Setup, \Oper)$, an adversary $\Adv$, and a number of rounds $r$.
    \begin{description}
        \item[Initialization:] 
        The challenger $\Challenger$ executes the setup algorithm to get $(\Dev_\init, \Key_\pub, \Key_\hid) \leftarrow \Setup(\SecPar, \Dev)$. $\Key_\pub$ is given to $\Adv$. The current state is set as $\Dev_\state \coloneqq \Dev_\init$.
        
        \item[Challenge:] 
         $\Challenger$ picks a random bit $b \pick \Set{0,1}$ and then executes the following steps for $r$ rounds with $\Adv$ ($i = [1..r]$):
        \begin{enumerate}
            \item \label{Multi-PD:item:patterns}
            The adversary $\Adv$ sends to $\Challenger$ two $\Lay$-patterns:
            $$\Pat_0 \coloneqq \Pat_\pub^1 \cup \Pat_\pub^2 \quad \text{and} \quad
            \Pat_1 \coloneqq \Pat_\pub^1 \cup \Pat_\hid,
            $$  
            where $(\Pat_\pub^1, \Pat_\pub^2, \Pat_\hid)$ satisfy the following requirements:
            \begin{enumerate}[label=(\alph*)]
                \item 
                $\Pat^1_\pub$ and $\Pat^2_\pub$ contain only public requests;
                \item
                $\Pat_\hid$ contains only hidden requests;
                \item \label{Multi-PD:item:empty-hidden-request}
                $\Pat_\pub^2$ must be $\varnothing$ if $\Pat_\hid$ is $\varnothing$;
                \item \label{Multi-PD:item:scheme-specific-requirement}
                $\Pat_\pub^1$ and $\Pat_\pub^2$ additionally satisfy some {\em scheme-specific} requirements $\Rule^1$ and $\Rule^2$ respectively;
            \end{enumerate}
            
            \item
            Based on the selected bit $b$, $\Challenger$ executes the request pattern $\Pat_b$ on the device, in an order of its choice, and updates the current device state as: 
           $$
            \Dev_\state \leftarrow \Oper(\Dev_\state, \Pat_{b}, \Key_\pub, \Key_\hid).
            $$
            
            \item \label[Step]{Multi-PD:item:states}
             $\Challenger$ sends $\Dev_\state$ and/or WOnly(OpTrace($\Lay, \Pat_{b_i}$)) to $\Adv$.
        \end{enumerate}
        
        \item[Output:] 
        Finally, $\Adv$ outputs a bit $b^*$. The game then terminates with the output defined as $\PD^{\Lay}_{\Sigma,\Adv}(\SecPar, r) \coloneqq (b == b^*)$.
    \end{description}
    \end{minipage}
    }
    \end{center}
    \caption{Security game for multi-snapshot, device and trace-oriented plausible deniable system. The game involves $r$ rounds to model both single and multi-snapshot adversaries.}
    \label{figure:MulPD}
\end{figure}

\para{Device and Trace-Oriented \Lay-Layer PD}.
The security for a PD scheme can now be formalized through the CPA-style game in \Cref{figure:MulPD}. This game is played between a coercive adversary $\Adv$ and a challenger $\Challenger$ running a PD scheme $\Sigma$. $\Adv$ only knows  $\Key_\pub$ (for the public data that the scheme is not trying to hide), but not $\Key_\hid$\footnote{Otherwise, there is nothing to protect.}. The game is played for $r$ rounds: when $r$ = 1 the game models a single-snapshot adversary, when $r$ = $\poly(\SecPar)$ it models a multi-snapshot adversary.

At each round $i = [1..r]$, $\Adv$ is allowed to send two patterns $\Pat_0$ and $\Pat_1$. $\Pat_0$ is the concatenation of two public parts $\Pat^1_\pub$ and $\Pat^2_\pub$,\footnote{Note that in the CPA game in \cite{chakraborti2017datalair}, $\Pat_0$ is also allowed to contain hidden requests. While this seems to make our definition weaker, we show in \Cref{appendix:game:equivalence} that the CPA game defined in \Cref{figure:MulPD} is actually equivalent to that in \cite{chakraborti2017datalair} in this aspect.} while $\Pat_1$ is the concatenation of  $\Pat^1_\pub$ and an arbitrary hidden request pattern $\Pat_\hid$ (up to some restrictions that will be discussed soon). The challenger executes $\Pat_{b}$ by picking public and hidden requests in an order of its choice. The challenger then sends back the snapshot of the device and/or the operation traces.

The adversary should not be able to tell which patterns are executed. More specifically, we define the advantage of the adversary in the game to be $Adv(\Adv)\coloneqq |\Pr[\PD^{\Lay}_{\Sigma,\Adv}(\SecPar, r) = 1] - 1/2|$. This captures the exact requirement of PD---the execution of hidden requests $\Pat_\hid$ now can be interpreted as some other public requests $\Pat^2_\pub$. Indeed, $\Adv$ cannot tell the difference by investigating the snapshots and/or operation traces.

As discussed in \Cref{sec:trace-oriented_PD}, the WOnly($\cdot$) function needs to be applied to screen out the $\Read$ operations from the traces before they are sent to $\Adv$. \Cref{lemma:tPD_wORAM} states that if a PD solution is secure in a setup where \Read instructions do not leave traces, it can be converted to a secure write-only ORAM. However, if a PD solution is provably secure even if \Read instructions {\em leave} traces on the storage device, then it can be converted to a full ORAM via an analog of \Cref{lemma:tPD_wORAM}. Thus, it will suffer from ORAMs' efficiency lower bound \cite{goldreich1996software,boyle2016there,larsen2018yes,weiss2018there,hubavcek2019stronger}. For example, HIVE \cite{blass2014toward} can be proven secure even if \Read leaves traces; indeed, it employs this actively by explaining a hidden access as a public \Read. Unsurprisingly, HIVE is constructed based on fully-secure ORAMs.

\subpara{The Adversary Requests}.
Note the requirements put on the adversary's request patterns. First, $\Pat_0$ and $\Pat_1$ must share the same $\Pat^1_\pub$, as otherwise $\Adv$ (with $\Key_\pub$) can always win the game by checking the public data in the received snapshot. For a similar reason, requirement \ref{Multi-PD:item:empty-hidden-request} (\Cref{figure:MulPD}) is also necessary.

An essential difference between this security game and previous ones lies in requirement \ref{Multi-PD:item:scheme-specific-requirement} (\Cref{figure:MulPD}). Ideally, a scheme should be secure against all PPT adversaries. However, this is usually not easy to achieve in practice. Instead, previous attempts proved the security of their solutions by making various additional assumptions on the adversary. In this paper we show that these assumptions can be viewed as requirements on the $\Pat^1_\pub$ and $\Pat^2_\pub$ part of the adversary's requests in the security game. Thus, the game in \Cref{figure:MulPD} generalizes them as two parameters $\Rule^1$ and $\Rule^2$. In \Cref{section:comparison} we show that by instantiating these two parameters properly, the game can capture the security requirements of all existing PD systems. Moreover, this approach provides a way to compare different PD solutions, where schemes with less restrictive $\Rule^1$ and $\Rule^2$ are preferable in terms of security. We choose not to provide restrictions for $\Rule^1$ and $\Rule^2$. Such guidelines are not possible nor useful since $\Rule^1$ and $\Rule^2$ are solution-dependent.

\begin{definition}[Device/Trace-Oriented \Lay-Layer \PD]
\label{definition:MulPD}
For a layer $\Lay$, a \sysname $\Sigma = (\Setup, \Oper)$ is device/trace-oriented \Lay-Layer \PD
if for any \PPT adversary $\Adv$ in the game of \Cref{figure:MulPD}, it holds that $Adv(\Adv) \le \negl(\SecPar)$, where $Adv(\Adv)\coloneqq |\Pr[\PD^{\Lay}_{\Sigma,\Adv}(\SecPar, r) = 1] - 1/2|$.
\end{definition}




\section{Comparison}
\label{section:comparison}
\begin{table}[!t]
\caption{Comparison of existing PD solutions. An empty circle signifies that the solution does not satisfy the property at the top, while a black circle denotes that the solution satisfies the property. A half-full circle in the Invisible column denotes that the respective solution (StegFS, INFUSE) tried to be invisible but did not completely succeed.}\vspace{1em}
\label{tab:hierarchy}
\renewcommand{\arraystretch}{1.5}
\centering
\begin{footnotesize}
    \begin{tabular}{l | C{30pt} | C{46pt} | C{42pt} | C{34pt} | C{40pt} | C{1.2cm} | C{28pt} | C{30pt} | C{1.5cm} | C{38pt} }
    \hline\hline
    {\bf Schemes} & {\bf Year} & {\bf Snapshot} & {\bf Security Type} & {\bf Layer} & {\bf I/O Perf.} \scriptsize{($\Pub$/$\Hid$)} & {\bf Space Util.}  &  {\bf Data Loss} & {\bf No Add'l. Space} & {\bf Invisible} & {\bf Device}\\
    \hline
    StegFS98 \cite{anderson1998steganographic}& 1998 & Sin & device & FS &  - & $\approx$15\% & $\mdlgblkcircle$ & $\mdlgblkcircle$ & $\mdlgwhtcircle$ & General\\
    \hline
    StegFS99 \cite{mcdonald1999stegfs}& 1999 & Sin & device & FS &  0.86/0.06 & - &  $\mdlgblkcircle$ & $\mdlgblkcircle$ & $\circlebottomhalfblack$ & General\\
    \hline
    StegFS03 \cite{pang2003stegfs}& 2003 & Sin & device & FS &  0.06 & $>$80\% & $\mdlgblkcircle$ & $\mdlgblkcircle$ & $\mdlgwhtcircle$ & General\\
    \hline
    TrueCrypt \cite{Truecrypt}& 2004 & Sin & device & BD &  - & 100\% & $\mdlgblkcircle$ & $\mdlgblkcircle$ & $\mdlgwhtcircle$ & General\\
    \hline
    MobiFlage \cite{skillen2013mobiflage}& 2013 & Sin & device & BD & 0.95 & 100\% & $\mdlgblkcircle$ & $\mdlgblkcircle$ & $\mdlgwhtcircle$ & General\\
    \hline
    MobiPluto \cite{10.1145/2818000.2818046}& 2015 & Sin & device & BD & - & 100\% & $\mdlgblkcircle$ & $\mdlgblkcircle$ & $\mdlgwhtcircle$ & General\\
    \hline
    DEFTL \cite{jia2017deftl}& 2017 & Sin & device & FTL &  - & 100\% & $\mdlgwhtcircle$ & $\mdlgblkcircle$ & $\mdlgwhtcircle$ & NAND flash\\
    \hline
    DEFY \cite{DBLP:conf/ndss/PetersGP15}& 2015 & Mul & device & FS & - & 100\% & $\mdlgblkcircle$ & $\mdlgwhtcircle$ & $\mdlgwhtcircle$ & NAND flash\\
    \hline
    MobiCeal \cite{chang2018mobiceal}& 2018 & Mul & device & BD &  0.78 & - & $\mdlgblkcircle$ & $\mdlgblkcircle$ & $\mdlgwhtcircle$ & General\\
    \hline
    INFUSE \cite{chen2020infuse}& 2020 & Mul & device & FS &  0.94/0.03 & $>$100\% & $\mdlgblkcircle$ & $\mdlgblkcircle$ & $\circlebottomhalfblack$ & Certain NAND flash\\
    \hline
    PEARL \cite{chen2020pearl}& 2021 & Mul & device & FTL &  0.6/0.15 & 80\% & $\mdlgblkcircle$ & $\mdlgblkcircle$ & $\mdlgwhtcircle$ & NAND flash\\
    \hline
    HIVE \cite{blass2014toward}& 2014 & Mul & trace & BD &  - & 50\% & $\mdlgwhtcircle$ & $\mdlgblkcircle$ & $\mdlgwhtcircle$ & General\\
    \hline
    HIVE-B \cite{blass2014toward}& 2014 & Mul & trace & BD &  0.004 & 50\% & $\mdlgwhtcircle$ & $\mdlgwhtcircle$ & $\mdlgwhtcircle$ & General\\
    \hline
    DataLair \cite{chakraborti2017datalair}& 2017 & Mul & trace & BD & 0.19/0.01 & 50\% & $\mdlgwhtcircle$ & $\mdlgwhtcircle$ & $\mdlgwhtcircle$ & General\\
    \hline
    ECD \cite{zuck2017preserving}& 2017 & Mul & trace & FTL &  $*$ & 52.5\% & $\mdlgblkcircle$ & $\mdlgwhtcircle$ & $\mdlgwhtcircle$ & NAND flash\\
    \hline
    PD-DM \cite{chen2019} & 2019 & Mul & trace & BD &  0.10/0.07 & $\approx$50\% & $\mdlgwhtcircle$ & $\mdlgblkcircle$ & $\mdlgwhtcircle$ & General\\
    \hline\hline
    \end{tabular}
\end{footnotesize}
\end{table}
In this section we compare the security and performance of existing PD schemes. We leverage the unified definition in \Cref{section:definition} to provide a framework for comparing the security of existing solutions. We further perform the comparison from a variety of aspects (summarized in \Cref{tab:hierarchy}), to provide a comprehensive understanding of these schemes.

\para{Security Metrics}.
The security guarantees of PD schemes are related to the assumptions made on adversaries, which can be captured by the unified definition. Specifically, the snapshot frequency and the type of security (listed in \Cref{tab:hierarchy}) categorize the scheme in coarse granularity, and the constraints $\Rule_1$ and $\Rule_2$ are used to characterize the power of adversaries in a finer way. Before presenting the constraints on each scheme, let us interpret the meaning of these constraints:

\begin{itemize}
\item
The ideal scheme should be secure against all PPT adversaries (corresponding to empty $\Rule_1$ and $\Rule_2$). No existing solutions achieve this level of security. $\Rule_1$ and $\Rule_2$ can be viewed as specifying a subset of all PPT adversaries against which a PD scheme is secure. Thus, they provide a criterion for security comparison: {\em the more constrictive, the fewer adversarial behaviors are ruled out, resulting in a more powerful adversary and a more secure scheme}.

\item 
The constraints also define under which conditions hidden operations can be executed safely. For example, the $\Rule^1_{\scriptscriptstyle{\sf DEFTL}}$ and $\Rule^2_{\scriptscriptstyle{\sf DEFTL}}$ for DEFTL below essentially say that the hidden operations can be performed with {\em any} public operation, as long as an $\mathtt{Unmount}$ is performed (together with the trigger post-processing), before the device is handed over to the coercive adversary. The constraints for other schemes can also be interpreted similarly. Thus, {\em the less constrictive the constraints, the more flexibility a scheme has in performing hidden operations}.
\end{itemize}

The following  interprets the security of existing schemes by specifying their corresponding constraints, and draw comparisons along the way.

\subsection{Single vs. Multi-snapshot Adversary}
\label{section:comparison:singlemulti}

To achieve single-snapshot security, existing solutions explore two major directions. The first direction is inspired by classical {\em steganography}, i.e., embedding relatively small messages within large cover-texts, such as adding imperceptible echoes at certain places in an audio recording \cite{petitcolas1999information}. Anderson \etal \cite{anderson1998steganographic} explored steganographic file systems and proposed two approaches for hiding data. The first approach defines the target file as the (password-derived) linear combination of a set of cover files. The second approach encrypts the target file using a block cipher with password-derived secret keys, and then stores it at the location determined by a cryptographic hash of the filename. In both approaches, an adversary without the correct password can get no information about whether the protected file ever exists. The latter approach was later implemented and optimized by McDonald \etal \cite{mcdonald1999stegfs} and Pang \etal \cite{pang2003stegfs}. Unfortunately, such approaches are not extremely space-effective, and come with potential data loss and high performance overheads. They are not suited for building modern systems handling large amounts of data at high speed.

The StegFS series \cite{anderson1998steganographic,mcdonald1999stegfs,pang2003stegfs} have the same model and share the same constraints in the unified definition:
\begin{itemize}
    \item $\Rule^1_{\scriptscriptstyle{\sf StegFS}}$: no restrictions ($\Pat^1_\pub$ can be any pattern); 
    \item $\Rule^2_{\scriptscriptstyle{\sf StegFS}}$: $\Pat_\pub^2$ must be an empty pattern.
\end{itemize}

The second direction \cite{Truecrypt,skillen2013mobiflage,jia2017deftl} handles PD at block-device level by designing disk encryption tools that help users embed ``hidden volumes'' (together with ``public volumes'') within the device (e.g., in the free space regions), while preventing adversaries from learning how many such volumes the device actually contains. Different keys are used to encrypt different volumes using randomized encryption indistinguishable from pseudo-random free space noise. Upon coercion, a user can provide the encryption keys for the public volumes, thus providing a plausible non-hidden use case for the disk. The adversary does not have any evidence for the existence of additional volumes.

\subpara{TrueCrypt \cite{Truecrypt}} successfully implemented this idea. It stores hidden volumes in the free space of public volumes. To hide their existence, TrueCrypt fills all free space with random data and encrypts the hidden data with a semantically secure encryption scheme that has pseudo-random ciphertexts. Upon coercion, the user can reveal the keys for the public volumes, and claim that the remaining space contains random free space. Rubberhose \cite{rubberhose}, MobiFlage \cite{skillen2013implementing} and DEFTL \cite{jia2017deftl} are implementations following  similar ideas targeted to different use cases (mobile devices, NAND flash). 

In the unified security definition, the constraints of TrueCrypt and MobiFlage are the following:
\begin{itemize}
    \item $\Rule^1_{\scriptscriptstyle{\sf Tc\&Mf}}$: no restrictions; 
    \item $\Rule^2_{\scriptscriptstyle{\sf Tc\&Mf}}$: $\Pat_\pub^2$ must be an empty pattern.
\end{itemize}

Further, DEFTL has the following constraints:
\begin{itemize}
	\item $\Rule^1_{\scriptscriptstyle{\sf DEFTL}}$: the last operation in $\Pat_\pub^1$ must be \texttt{Unmount};
	\item $\Rule^2_{\scriptscriptstyle{\sf DEFTL}}$: $\Pat_\pub^2$ must be an empty pattern.
\end{itemize}

\para{Highlights}:
Single-snapshot security can be achieved with low overheads and high performance. Although these schemes are designed in different storage layers (\FS vs \BD), they share the same restrictions on the adversaries' choice of patterns, thus achieving identical security guarantee. However, all the aforementioned schemes fail to protect against multi-snapshot adversaries. For example, when TrueCrypt writes hidden data, the device ``free space'' changes unexplainably. When observed by a multi-snapshot adversary this cannot be plausibly explained away. After all, why did the disk free area change without corresponding substantial changes to the public data?

Thus, to protect against multi-snapshot adversaries, one needs to {\em hide not only the existence of hidden data, but also associated access patterns.}

%
%

Given this insight, progress in this area centers mainly around mechanisms that can consistently explain updates to both the public and hidden data across multiple snapshots. Currently, three major approaches exist: (i) using oblivious RAM mechanisms (\Cref{sec:hide_pattern_ORAM}), (ii) using canonical forms (\Cref{section:canonical-forms}) and (iii) relying on device/deployment-specific properties (\Cref{sec:fine-tuned-solution}). The remainder of this section will explore these approaches, aiming to understand the fundamentals and distill insights to guide future designs.

\subsection{ORAM-Based PD Schemes}
\label{sec:hide_pattern_ORAM}

Multi-snapshot secure PD requires mechanisms that hide users' access patterns to hidden data. ORAMs \cite{goldreich1996software} are natural tools for this task. 

\noindent{\em ORAMs}.
Roughly, an ORAM ensures a database-hosting server cannot determine which database (the ``RAM'') entries are accessed by one of its client. Access patterns of any same-length access sequences are designed to be indistinguishable. As a simple example, a (highly inefficient yet secure) ORAM (with $O(n)$ asymptotic complexity per access) can be constructed by filling the database with randomized encrypted data; To access one of the elements, the client reads the entire database, re-encrypts it and writes it back to the server. More efficient solutions exist that enable complexities much lower than $O(n)$ \cite{stefanov2011towards,stefanov2013path,DBLP:conf/ndss/ChakrabortiACMR19,DBLP:conf/ndss/ChakrabortiS19,DBLP:journals/popets/ChakrabortiS20}.  An exhaustive treatment is out of scope here. The following summarizes how ORAMs were employed to obtain PD solutions. 

\subpara{HIVE \cite{blass2014toward}} was the first to deploy ORAMs. It introduced {\it hidden volume encryption}. The main idea was similar to a previous work \cite{Truecrypt}, i.e., to divide the storage into a public volume and a hidden volume\footnote{The original HIVE scheme supports multiple volumes. W.l.o.g., only the two-volume case is considered here (for simplicity).}, each volume being accessed using an ORAM mechanism. 
Additionally, for every access to a volume (either public or hidden), the system also executes {\it dummy} accesses to the other volume. Since ORAM accesses are indistinguishable from each other (whether dummy or not), adversaries cannot tell the difference between 1) accesses to the public volume and 2) accesses to the hidden volume, {\em which satisfies the exact requirement of the CPA game for PD}.

Moreover, HIVE leveraged the observation that \Read operations are not visible to adversaries, since such operations do not leave any discernible traces (\Cref{section:model:system}). Thus, it is sufficient to use write-only ORAM schemes (see \Cref{def:wORAM} in \Cref{sec:wORAM_from_PDs}) that only hide \Write operations. HIVE \cite{blass2014toward} designed a specific write-only ORAM with a small stash of pending blocks in memory, i.e., where blocks are stored to be written later when a free block becomes available. The write-only ORAM stash can also behave as a queue for caching hidden data, and hidden volume accesses can be performed together with existing (if any) public volume accesses to minimize the need for additional dummy accesses. In this case, the adversary cannot tell the difference between 1) accesses to the public volume only and 2) accesses to both public and hidden volumes. The only associated requirement now becomes the need for enough plausible public accesses to pair with the hidden data in the stash when written to disk. 

In the unified definition, HIVE has the following constraints:
\begin{itemize}
	\item $\Rule^1_{\scriptscriptstyle{\sf HIVE}}$: $\Pat_\pub^1$ and $\Pat_\hid$ must be of equal length;
	\item $\Rule^2_{\scriptscriptstyle{\sf HIVE}}$: $\Pat_\pub^2$ must be an empty pattern.
\end{itemize}

\subpara{HIVE-B \cite{blass2014toward}} is another PD scheme proposed in the same paper as HIVE. It provides the same security guarantee as HIVE, but with different constraints:
\begin{itemize}
\item 
    $\Rule^1_{\scriptscriptstyle{\sf HIVEB}}:$ no restrictions;
\item 
    $\Rule^2_{\scriptscriptstyle{\sf HIVEB}}:$ $\Pat_\pub^2$ and $\Pat_\hid$ must be of equal length (i.e.~containing the same number of requests).
\end{itemize}

\subpara{DataLair \cite{chakraborti2017datalair}} extends these ideas and observes that operations on public data do not need to be hidden since they are anyway public. In fact, revealing operations on public data reinforces deniability as it
shows plausible non-hidden device use. Therefore, DataLair only uses $\wORAM$s for the hidden volumes, while allowing public data to be accessed (almost) directly without any oblivious access mechanism. Moreover, it designs a specific throughput-optimized \wORAM. Following the strategy of HIVE, it pairs the operations on hidden data with those on public data, and ensures that such executions are indistinguishable from the operations on public data alone. Compared with its predecessors, DataLair accelerates public operations by two orders of magnitude, and also speeds up hidden operations.

In the unified definition, DataLair introduces a parameter $\phi$ in the constraints:
\begin{itemize}
	\item $\Rule^1_{\scriptscriptstyle{\sf DataLair}}$: $\Pat_\pub^1$ should contain at least $\phi \times k$ public \Write operations where $k$ is the length of $\Pat_\hid$ and $\phi$ is a pre-defined parameter;
	\item $\Rule^2_{\scriptscriptstyle{\sf DataLair}}$: $\Pat_\pub^2$ must be an empty pattern.
\end{itemize}

We note that DataLair's $\Rule^1$ constraints are stronger than, e.g., StegFS. However, this does not imply that StegFS provides stronger security because the security game models the adversary's power also through the $r$ parameter: StegFS is designed for single-snapshot, while DataLair is designed for multi-snapshot adversaries.

\subpara{MobiCeal \cite{chang2018mobiceal}} implements PD at the \BD layer, and supports a broad deployment of any block-based file systems for mobile devices. MobiCeal improves performance by replacing wORAMs with {\it dummy} write operations coupled to public writes. In the unified definition, MobiCeal has the following constraints, where $f \in (0, 1)$ is a random number and $\lambda$ is a rate parameter:
\begin{itemize}
	\item $\Rule^1_{\scriptscriptstyle{\sf MobiCeal}}$: For each public write in $\Pat_\pub^1$, also perform a dummy write with a certain probability. The dummy write contributes $m$ dummy block writes, where $m$ is chosen according to an exponential distribution, $m =  \lfloor -(\ln(1 - f))/\lambda \rfloor$.
	%
	\item $\Rule^2_{\scriptscriptstyle{\sf MobiCeal}}$: $\Pat_\pub^2$ must be an empty pattern.
\end{itemize}

\para{Where to Write}.
An important factor affecting both the security and efficiency of ORAM-based PD approaches \cite{blass2014toward,chakraborti2017datalair} is {\em free-block allocation} (FBA), i.e.~the mechanism to keep track of free blocks to store new incoming data. 

Note that HIVE uses separate ORAMs on public and hidden storage spaces, and DataLair uses ORAM only for hidden space. A naive approach would be to have separate FBA mechanisms for public and hidden spaces.
Unfortunately, this can lead to storage capacity waste, as the hidden space must be allocated even if it is never used. Instead, a better solution uses a ``global'' FBA algorithm across all the storage space. In this case, both the public and the hidden volume can be of the same logical size as the underlying partition, and use all the available space for either hidden or public data.

However, this turns to be a delicate task due to the existence of hidden data. On the one hand, the FBA should avoid overwriting existing hidden data; On the other hand, such avoidance should be strategically hidden to not raise doubts from the adversary about the existence of hidden data.

Moreover, since it is in the data path, FBA must be efficient. Significant amount of work has been devoted \cite{blass2014toward,chakraborti2017datalair} to the design of FBA algorithms that meet the above criteria. The reader is referred to the original papers for details. 

%
%

\subsection{Replacing Randomization with Canonical Forms}
\label{section:canonical-forms}

ORAMs are used in PD designs because they can hide both the locations and contents of each access, mostly via inherently high-overhead randomization.  Yet, randomization is not really necessary to achieve plausible deniability\cite{chen2019}. Simple canonical forms -- e.g., such as used in log-structured file systems \cite{douglis1989log} always writing data sequentially, treating the logical address space as a circular buffer -- may be enough to decouple the user's logical from physical access patterns.

Since canonical forms ensure pre-defined physical device write traces, an adversary is prevented from inferring the logical layer access patterns, of which the traces are independent of. 

Further, importantly, an advantage of certain canonical forms (e.g., sequential) is the ability to retain data locality and thus result in significantly higher efficiency than randomization-based ORAM approaches. 

\subpara{PD-DM \cite{chen2019}} is the first work that {\em explicitly} notes the above idea. Its design ensures that all writes to the physical device are located at sequentially increasing physical addresses, similar to \texttt{Append} operations in log-structure file systems. PD-DM stipulates that whenever a public data record $D_\pub$ is written on the device, an additional random string $R$ (the ``payload'') is written immediately in the immediately adjacent next block. To store hidden data, PD-DM will first encrypt it (indistinguishably from random) and then write it as  the payload of some public data write $D_\pub$. In this case, device snapshots look like the following:
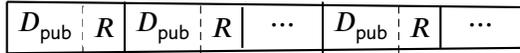
\begin{center}
\begin{tikzpicture}[
long/.style = {rectangle, 
                  thin,
                  fill=white, 
                  align=center,
                  minimum height=15pt,
                  minimum width=30pt},
large/.style = {rectangle, 
                  thin,
                  fill=white, 
                  align=center,
                  minimum height=15pt,
                  minimum width=26pt},
small/.style = {rectangle, 
                  thin,
                  fill=white, 
                  align=center,
                  minimum height=15pt,
                  minimum width=15pt},
]
\node [large] (l1) {$D_\pub$};
\draw [thick] (l1.south west) -- (l1.north west);
\draw [densely dashed] (l1.south east) -- (l1.north east);
\node [small, right = 0pt of l1] (s1) {$R$};

\node [large, right = 0pt of s1] (l2) {$D_\pub$};
\draw [thick] (l2.south west) -- (l2.north west);
\draw [densely dashed] (l2.south east) -- (l2.north east);
\node [small, right = 0pt of l2] (s2) {$R$};

\node [long, right = 0pt of s2] (long) {$\cdots$};
\draw [thick] (long.south west) -- (long.north west);

\node [large, right = 0pt of long] (l3) {$D_\pub$};
\draw [thick] (l3.south west) -- (l3.north west);
\draw [densely dashed] (l3.south east) -- (l3.north east);
\node [small, right = 0pt of l3] (s3) {$R$};

\node [long, right = 0pt of s3] (long2) {$\cdots$};
\draw [thick] (long2.south west) -- (long2.north west);

\draw [thick] (l1.north west) -- (long2.north east);
\draw [thick] (l1.south west) -- (long2.south east);
\end{tikzpicture}
\end{center}

Since the encrypted hidden data looks indistinguishable from the random ``payload'' of public writings, adversaries are unable to distinguish whether any hidden data exists or not. 

In terms of the unified definition constraints, PD-DM has similar constraints to DataLair \cite{chakraborti2017datalair}. However, while DataLair requires a fixed-value parameter $\phi$, in PD-DM the value of $\phi$ is system specific.

%

\subpara{ECD \cite{zuck2017preserving}} works in a related manner. The idea is to partition the device into a public and a hidden volume. The public volume is managed by the system in the standard way, independent of the hidden one. The hidden volume is divided into equal-size sequential {\em segments}, denoted as $\Set{s_1 , s_2, \ldots, s_N}$. Each $s_i$ contains some free blocks containing random strings, while other blocks may be occupied by encrypted hidden data. ECD keeps moving data from $s_{i-1}$ to $s_i$ at a predetermined rate. During this procedure, any free blocks will be re-randomized, while existing hidden data blocks will be re-encrypted. To a polynomial adversary this looks like all of $s_i$ is re-randomized. New hidden data is written encrypted into a free block during the migration from $s_{i-1}$ to $s_i$. Overall this can be viewed as creating an artificial canonical form on the hidden volume, where segment $s_i$ is periodically overwritten by its immediate predecessor $s_{i-1}$. 
In the unified model, ECD has the following constraints:
\begin{itemize}
	\item $\Rule^1_{\scriptscriptstyle{\sf ECD}}$: no restrictions;
	\item $\Rule^2_{\scriptscriptstyle{\sf ECD}}$: $\Pat_\pub^2$ must be an empty pattern.
\end{itemize}

ECD has the least restrictive rules compared with other schemes. This is because it employs a regular system behavior which periodically modifies the state of the device to cover up the hidden operation; Such a periodic system update is general enough to cover any type of hidden operations. In contrast, other schemes use public operations to cover up hidden operations, which naturally introduces constraints to ensure that the public and hidden operations are ``paired'' properly.

We conclude that HIVE, DataLair and PD-DM provide multi-snapshot security at the $\BD$ layer, with comparable constraints. They all achieve the stronger trace-oriented security.


%
%

\subsection{Device-Specific Mechanisms}
\label{sec:fine-tuned-solution}

In pursuit of performance, a recent line of work emerged building plausible deniability guarantees on device-specific properties, e.g., electric charge levels in flash memories. This enjoys certain advantages over previous work: 1) it may avoid heavy machinery (e.g., ORAMs) and may lead to lightweight solutions; 2) resulting mechanisms may be closer or even native to the underlying device, allowing for higher performance and better plausibility. 

This approach was often implicit in the literature. This section seeks to sublimate the essence of such constructions in a unified perspective. First, consider several existing constructions. 

\subpara{INFUSE \cite{chen2020infuse}} builds a PD scheme in the flash \FTL layer. The main idea is to modulate additional information in charges and voltage levels of individual NAND cells, the minimal storage unit for NAND. A cell can hold one (SLC, single-level cell) or more (MLC, multiple-level cell) data bits. Bits are encoded and decoded by using a programmable {\em threshold voltage} $V_{\mathsf{th}}$ and a predefined {\em reference voltage} $V_{\mathsf{r}}$. For example, an SLC cell with threshold voltage $V_{\mathsf{th}} = 3V$ will be interpreted as a logical ``1'' when the reference voltage level is $V_{\mathsf{r}} = 3.5V$, and as a ``0'' if either (i) the reference voltage level drops below $V_{\mathsf{r}} = 2.5V$ or (ii) the threshold voltage is increased to e.g., $V_{\mathsf{th}} = 4V$. MLC work similarly, with multiple levels to encode multiple values. 

Some recent flash controllers are able to operate the same cell in both SLC and MLC mode \cite{lee2009flexfs}. This provides an opportunity to hide bits. Multiple bits can be stored in a particular cell using an ``MLC-style'' encoding but on inspection the system can claim that the cell is in SLC mode and provide only a single bit. Care needs to be taken to ensure device-wide indistinguishability between sets of cells in either SLC or MLC mode. This constitutes the core idea of INFUSE. 

Under the (mostly empirical) assumption that an adversary cannot distinguish which cells are used in which mode or whether there are any inconsistencies in the distribution of SLC vs MLC cells, this scheme provides significant speedups. Public data operations are orders of magnitude faster than existing multi-snapshot resilient PD systems, and only 15\% slower than a standard non-PD baseline and hidden data operations perform comparably to the-state-of-the-art PD systems.

In the unified definition, INFUSE has the following constraints:
\begin{itemize}
	\item $\Rule^1_{\scriptscriptstyle{\sf INFUSE}}$: the last operation in $\Pat_\pub^1$ must be \texttt{Unmount};
	\item $\Rule^2_{\scriptscriptstyle{\sf INFUSE}}$: $\Pat_\pub^2$ must be an empty pattern.
\end{itemize}

\subpara{PEARL \cite{chen2020pearl}} is also operating in the \FTL layer, but relies on a new smart {\em write-once memory} (WOM) encoding that does not require custom voltage programming. 

Unfortunately, once written to, a NAND flash cell cannot be reprogrammed before an {$\Erase$} of its containing block. Further, NAND flash is reliable only for a limited number of {$\Erase$} cycles. This can severely limit device lifespan. Complex wear leveling algorithms are deployed to ``even out'' wear and maximize lifespan. 

WOM codes \cite{rivest1982reuse} have been proposed to further optimize this wear. They use an important property of NAND flash: previously-unwritten-to cells can be written to even if they are in pages that have been written to before. 
WOM codes encode with enough redundancy (e.g.~using 3 cells to store 2 bits) to allow {\em multiple} writes to the same page (i.e., with different data each time) without requiring an {$\Erase$}. 

At a high level, in the subsequent (e.g., second) \Write, the idea is to modify only the bits that have not been written-to in the first \Write. A well-designed encoding allows the second logical \Write to be encoded in the resulting physical state with no ambiguity.

For example, consider the case of an encoding with 2-bit logical data records encoded onto 3 physical bits. For each 2-bit logical record $s\in \bits^2$ the encoding defines two possible physical 3-bit configurations $\Enc_1(s)$ and $\Enc_2(s)$.  When logical record $s$ is stored {\em for the first time}, $\Enc_1(s)$ is stored physically. If the logical record $s$ needs to be replaced with a new value $s'$ (at the same location) writing simply converts the physical value $\Enc_1(s)$ to $\Enc_2(s')$. The WOM encoding is designed unambiguously and in such a way that any such conversion does not require overwriting an existing written-to cell. Since such a code allows two \Write operations per erasure, it is called a 2-write WOM code\footnote{W.l.o.g., for simplicity this work focuses on 2-write WOM codes. There exist $k$-write WOM codes that admit $k$ writings per erasure~\cite{godlewski1987wom,yaakobi2012codes,shpilka2013new}. }.

PEARL \cite{chen2020pearl} hides information by {\em modulating the written public data according to the data to be hidden}. To this end, it re-purposes WOM codes. When public data is written, the codeword is chosen based on the bits of the data that need to be hidden. This enables PEARL to surreptitiously hide information even in the presence of a powerful multi-snapshot adversary. The end-result is device state that is indistinguishable from the case of a device that was simply writing data multiple times using a WOM code. Much care needs to be taken in the design of the specific WOM code to not introduce device-wide bias. Overall however, the fact that WOM codes are widely deployed on NAND flash further strengthens plausibility. Most importantly, the resulting performance is comparable to the non-PD baseline on real-world workloads!

In the unified definition, PEARL has the following constraints:
\begin{itemize}
	\item $\Rule^1_{\scriptscriptstyle{\sf PEARL}}$: the last operation in $\Pat_\pub^1$ must be \texttt{Unmount};
	\item $\Rule^2_{\scriptscriptstyle{\sf PEARL}}$: $\Pat_\pub^2$ needs to generate $k$ 1st invalid pages where $k$ is the length of $\Pat_\hid$.
\end{itemize}

A pattern that can generate a 1st invalid page can be: 1) one public \Write followed by a public $\mathtt{Delete}$ to the same page; 2) one public \Write which has a correspondingly public \Write in $\Pat_\pub^1$. 

\subpara{DEFY \cite{DBLP:conf/ndss/PetersGP15}} is a log-structured \FS for NAND flash that offers PD with a newly proposed secure deletion technology. It is based on WhisperYAFFS~\cite{whisper}, a log structured \FS which provides full disk encryption for flash.  Log-structured {\FS}es have two relevant properties:
\begin{enumerate}
\item
Data (e.g., files, directories, links) and metadata are stored sequentially within the logical address space, and any access to data (including \Read) can cause the update of its corresponding metadata;
\item
Updates/deletes of data and metadata will not cause an actual deletion. Instead, a new address will be assigned to the updated version, and the old data/metadata is just marked as old; Subsequent garbage collection handles it.
\end{enumerate} 

DEFY achieves PD by exploiting the above properties in the following way. In DEFY, modifications to hidden data will cause the allocation of new records. Such allocations can be claimed as the results of metadata updating due to public \Read/\Write, since such updating can also lead to the assignment of new records. Once these records (for hidden data) become obsolete (i.e.~succeeded by newly allocated records), the system can claim that they were due to public accesses and are now securely deleted. Due to the irreversibility of secure deletion, the adversary has no choice but to believe that these records were due to public accesses. The system thus denies the existence of hidden data successfully.

DEFY enjoys impressive efficiency. \Read operation can be as fast as the Linux EXT4 file system. Further, in the unified definition, DEFY offers PD conditioned on constraints:
\begin{itemize}
        \item $\Rule^1_{\scriptscriptstyle{\sf DEFY}}$: $\Pat_\pub^1$ should contain some public operations, and the last operation in $\Pat_\pub^1$ must be \texttt{Unmount};
        \item $\Rule^2_{\scriptscriptstyle{\sf DEFY}}$: $\Pat_\pub^2$ should contain public accesses that generate enough deleted pages to cover accesses in $\Pat_\hid$.
\end{itemize}

Unfortunately DEFY is not secure, and it can be compromised in a few attempts to exhaust the writing capacity \cite{jia2017deftl}. Also, as hidden data are stored masqueraded as securely deleted obsolete (public) data, to maintain plausibility, the space occupied by them must be plausibly and frequently enough overwritten by public data. This results in data loss.

In addition, DEFY assumes the existence of a special, ``tag storage area'' on the device that is hidden from the adversary. This breaks security against multi-snapshot adversaries. Thus, the security provided by DEFY is considered weaker than the schemes that do not assume the existence of such a hidden area.


%

\para{In summary}, these schemes are optimized for specific deployment cases. INFUSE and PEARL exploit voltage variation and properties of WOM codes respectively, to encode hidden data together with public data at the same locations. DEFY plausibly encodes hidden data as securely-deleted obsolete public data. The resulting solutions are specific to the underlying devices but achieve performance comparable to the non-PD baseline on real-world workloads. Such efficiency is clearly out of the reach of ORAM-based solutions. Importantly, WOM code-based schemes provide an unusually favorable combination of strong security and high performance.

%
%


\subsection{Access Pattern Hiding Techniques}
\label{sec:ana_access_pattern}


As mentioned earlier, a key point of PD schemes is to conceal access patterns to hidden data. In order to hide the existence of hidden data, a PD scheme should prevent adversaries from learning not only {\it which hidden access happens}, but also {\it how many hidden accesses happen}. This is in contrast to ORAMs where only that the access patterns of logical requests are not revealed, while the number of accesses can be public.

To hide {\em which hidden access happens}, existing PD schemes leverage one of the following two strategies: 1) randomizing the write trace on physical devices; 2) enforcing the write trace to follow certain canonical form (the commonly used one is log-structure). HIVE, DataLair and MobiCeal follow the first strategy; PD-DM, ECD, DEFY, INFUSE and PEARL follow the second strategy. The last 3 schemes are designed for NAND flash devices, where the device is written sequentially by default. To hide the {\it number of hidden accesses}, the first strategy is to make the change of device state due to a hidden access to be indistinguishable from that of some non-hidden accesses. Thus, any changes on the devices can be attributed to certain public accesses, and the number of hidden accesses can be claimed as 0. 
Note that the public accesses used to ``explain'' hidden accesses do not need to happen in reality.  The PD schemes that use this strategy are HIVE, DEFY, PEARL and ECD. HIVE explains a hidden access as a public \Read. DEFY and PEARL explains it either as a public \Write or $\mathtt{Delete}$, while ECD explains a hidden access as a system behavior that happens at a pre-defined rate.

Another strategy to hide the number of hidden accesses is to ``pair'' hidden accesses with some public accesses and ensure that the write trace of the public accesses alone is indistinguishable from that of both the public accesses and hidden accesses. As a result, for an adversary, only public accesses happen. Examples include HIVE-B, DataLair, PD-DM, Mobiceal and INFUSE.


\subsection{Performance Metrics}


\Cref{tab:hierarchy} also looks at existing solutions from a performance standpoint.

\para{I/O Performance}.
PD comes with both throughput and space overheads. Some schemes report performance separately for public operations and hidden operations -- shown in \Cref{tab:hierarchy} in the form of ``$x$/$y$''. It means that the public throughput is $x$ times of the non-PD baseline, and the hidden throughput is $y$ times of the baseline. Some other schemes reported only one overall performance number, and some schemes did not provide any explicit performance number or even have not been evaluated at all since they are designed in theory and no implementation is completed (shown as ``-'' in \Cref{tab:hierarchy}). 
ECD is a special case whose performance (marked with ``$*$'' in the table) depends on a system parameter. Recall that ECD covers up hidden operations by periodically updating the device state at a prefixed rate $r$. Thus, the number of hidden operations that the system is able to perform is determined by the updating rate $r$, rather than the public operations.

\para{Space Utilization}.
The column ``{\bf Space Util.}'' shows how efficiently the storage capacity of physical devices can be exploited by each PD scheme. It is computed as the ratio {\em between} the max size of data (both public and hidden) that can be stored in one device {\em and} the total capacity of the storage device as a metric for space utilization in \Cref{tab:hierarchy} (space required by meta-data is excluded for simplicity). Note that INFUSE enjoys a space utilization larger than 100\%. That is because INFUSE encodes hidden bits at physical storage cells that already contain some public data (see \Cref{sec:fine-tuned-solution}).

\para{Additional Safe Space, Data Loss}.
As discussed earlier, some PD solutions assume the existence of an area on the devices that remains hidden from adversaries (e.g., the TSA block in DEFY, or the stash in HIVE). Some schemes suffer from data loss, i.e., the hidden data may be overwritten (maybe by public data) in some use cases. \Cref{tab:hierarchy} also lists these caveats for each scheme.


\section{Key Insights}
\label{sec:key-insights}


PD solutions deployed in a layer do not necessarily ensure PD for the entire system (\Cref{section:depedence-on-layers}). However, we have shown that trace-oriented PD implies the standard PD security, and trace-oriented PD secure mechanisms can provide PD for the entire system. More specifically, since traces at a layer are converted through a PPT algorithm into traces at at lower layer (\Cref{sec:trace-oriented_PD}), indistinguishability of traces at a layer implies indistinguishability of traces and snapshots at any lower-layers. This addresses the issue of the SSD/FTL example in \Cref{section:model:adversary}: if the PD solution is BD layer trace-oriented secure, it achieves plausible deniability even though the SSD has an FTL layer below.

A key point of multi-snapshot resilient PD systems lies in hiding access patterns to hidden data (\Cref{section:comparison:singlemulti}). ORAMs have been used to build PD schemes that hide access patterns (\Cref{sec:hide_pattern_ORAM}). However, ORAM-based solutions are inefficient due to the inherently heavy randomization machinery. Further, ORAMs require carefully-designed free-block allocation algorithms.

To improve performance, existing work has explored two directions, canonical form and device-specific solutions. Canonical form-based PD solutions can hide user access patterns, and significantly increase throughput (\Cref{section:canonical-forms}). In particular, sequential approaches can preserve data locality and make good use of locality-optimized systems deploying caching and read-ahead mechanisms. Further, lightweight, device-specific PD solutions have been developed, that exploit specific devices and deployment settings to achieve efficiency comparable to the non-PD baselines, that does not always come at the expense of strong security (\Cref{sec:fine-tuned-solution}).

We also note that several PD solutions impose $\Pat^2_\pub$ to be empty. This is because to conceal hidden accesses, some PD schemes make the device state associated with a hidden access be indistinguishable from that of some non-hidden accesses.

\section{Future Directions}
\label{sec:future-direction}

We leverage these insights to propose several promising directions for future work.


\subsection{Trace-Oriented Security}


Most existing PD solutions exploit the possibility that different traces may result in the same snapshot, which allows users to interpret hidden operations as public ones. However, such an advantage is lost in the trace-oriented setting, as the adversary obtains {\em actual} traces. Thus, it is necessary to remove any clues of the user's operations from the traces. \Cref{lemma:eq:woram-tpd} establishes the equivalence of trace-oriented PD to $\wORAM$s. Thus, this goal is hard to achieve without relying on {\wORAM}s.


Established lower-bounds for ORAMs can be viewed as a signal of the inefficiency of $\wORAM$s, which translates into clues to the inefficiency of robust PD. More specifically, strongly-secure PD solutions feature inherent fundamental efficiency limits, and achieving efficient PD requires layer-dependency.

However, future progress on \wORAM lower-bounds will also apply to trace-oriented PDs. Importantly, there is no {\em established} lower-bound for $\wORAM$s yet. Thus, the optimistic interpretation of \Cref{lemma:eq:woram-tpd} encourages us to seek efficient trace-oriented PDs. Given that all exiting trace-oriented PD solutions are built on top of $\wORAM$s, it will be interesting to have constructions that do not make {\em explicit} use of $\wORAM$s. Such constructions may circumvent the ORAM lower-bound (if it turns out it applies to $\wORAM$s). Thus, while it is challenging to build schemes achieving both trace-based security and good efficiency, this also yields the following insight: Instead of striving for trace-based PD, a more promising direction may be to take the approach illustrated in \Cref{sec:fine-tuned-solution}, and design PDs directly for the ``right'' layer.

More specifically, a careful selection of the layer at which the PD solution is implemented, if secure for traces from lower layers, may provide both trace-oriented security and efficiency. For example, for an HDD device whose block device manager does not shuffle the FS layout, an efficient FS-layer PD solution may be a better option than a trace-oriented PD solution. For an SSD device, or an HDD with a block device manager that shuffles the FS layout, an efficient PD solution may be implemented at the BD or FTL layers.


\subsection{Invisible PD}


Typically, PD systems only intend to hide the existence of hidden data, not the fact that {\it the system in use is PD}. However, the deployment of a PD system already raises suspicion about the existence of sensitive data. A similar issue also exists for deniable encryption \cite{canetti1997deniable,DBLP:conf/crypto/CanettiPP20}.

To equip the user with more credibility in the face of coercive authorities, future work may focus on {\it invisible} PD schemes that hide not only contents but also the evidence that the system is being used to hide data. This can be done by, e.g., making the scheme look indistinguishable to a off-the-shelf storage system. For instance, as shown in the ``invisible'' column of \Cref{tab:hierarchy}, StegFS \cite{mcdonald1999stegfs} was designed to be indistinguishable to EXT2. However, StegFS \cite{mcdonald1999stegfs} needs to also maintain a bitmap. Future efforts may look into making this scheme fully indistinguishable by removing the bitmap, while not compromising security.

INFUSE \cite{chen2020infuse} was designed to be indistinguishable from YAFFS \cite{yaffs}. However, INFUSE has a limited capacity for hidden data: If too much hidden data is stored, the distribution of cell voltages may become suspicious. Further, INFUSE requires the firmware support which allows precise manipulation on flash cell voltages. However, current NAND flash chips do not have the corresponding interface admitting such manipulations.

A promising direction is work on WOM codes~\cite{chen2020pearl} (\Cref{sec:fine-tuned-solution}), where information is surreptitiously hidden in the WOM codes of public data. While WOM codes are widely deployed on NAND flash, making it possible to deny the use of a PD solution, PEARL is based on customized ``PD-friendly'' WOM codes. Nevertheless, PEARL \cite{chen2020pearl} suggests that WOM codes have great potential for efficient PD constructions. Future research may focus on finding other PD-friendly WOM codes with improved efficiency, and further our understanding of PD-friendly WOM codes, e.g., proving necessary and efficient conditions for WOM codes to be PD-suitable, and lower-bounds on code rates for such codes.

\subsection{Explore Adversary Model Changes}

As discussed in \Cref{section:single-vs-multi}, designing PD schemes secure against multi-snapshot adversaries is challenging. Existing solutions are still too slow. To design a new PD scheme against multi-snapshot adversaries, one can either come up with a new strategy to hide the number of hidden accesses, or a new strategy to hide which hidden access happens, and then combine it with some of the exiting strategies listed in \Cref{sec:ana_access_pattern}. Finding new strategies for hiding the number of hidden accesses seems more promising as there could be different ways to interpret the disk changes resulting from hidden accesses.

A further promising direction is to design solutions secure against more realistic, bounded adversaries. Examples worth exploring include (lower) bounds on the number of operations that the user needs to perform between adversary-captured snapshots, or the total number of snapshots that an adversary can capture.

We also note however that assumptions A2 and A3 (\Cref{section:model:adversary}) underestimate the power of realistic adversaries, who can perform attacks that include cold boot attacks, access swap files and core dumps. Real-time access to, e.g., caches, allows inference of some \Read operations. Unfortunately, existing work ignores caches. Extending deniability to other parts of the system stack represents an interesting future direction. For instance, future work may treat caches and the DRAM as another layer in the storage hierarchy. We note however that a PD solution that is provably secure when \Read instructions leave traces on the storage device, can be converted to a full ORAM via an analog of \Cref{lemma:tPD_wORAM}, thus will suffer from ORAMs' efficiency lower bound \cite{goldreich1996software,boyle2016there,larsen2018yes,weiss2018there,hubavcek2019stronger} (\Cref{section:definition}).


\subsection{Synthetic Operations}


Existing PD schemes try to match hidden operations to public ones. This makes hidden operations rather passive: to perform a hidden operation, the system has to wait until the occurrence of the related public operation. It also restricts the types of allowed hidden operations.

Instead, an {\it active} approach is to let the system generate synthetic public operations whenever the user wants to perform hidden operations. Existing AI/ML solutions, e.g.,  variational autoencoders \cite{KW13} and generative adversarial networks \cite{GPAMXWOCB14}, trained on large sets of real-user operations, may be used to generate synthetic public operations that are difficult to distinguish from real public operations.


\section{Conclusion}


Plausible deniability can provide strong privacy guarantees that impacts millions of users in a world increasingly encroaching on encryption and personal privacy. Yet, building secure plausibly deniable efficient systems is far from trivial. This work systematizes existing knowledge for researchers and practitioners alike aiming to understand, deploy, or design plausible deniability systems. We believe plausible deniability to be an important property on the cusp of efficient mainstream practicality. This work is meant as a concise yet reasonably-complete guide on this journey.

\section{Acknowledgements}


We thank the shepherd, Diogo Barradas, and the anonymous reviewers for their feedback. This work has been supported by the National Science Foundation (award 2052951) and the Office of Naval Research (award N000142112407).

\bibliographystyle{plain}
\bibliography{myBib}
\addcontentsline{toc}{section}{References}

\appendix

\input{appendices}

\end{document}

%% file: macros.tex
\usepackage{comment}
\usepackage[normalem]{ulem}
\newcommand{\dlt}{\bgroup\markoverwith{\textcolor{red}{\rule[0.5ex]{1pt}{0.9pt}}}\ULon}

\newcommand{\red}[1]{{\color{red} #1}}

\usepackage{pifont}
\newtcolorbox[auto counter]{CommentBox}[2][]
{
  breakable,
  colframe = gray!10,
  colback  = gray!2,
  coltitle = Purple, 
  coltext  = Purple, 
  title    = #2,
  arc=0pt,
  outer arc=0pt,
  boxrule=1.5pt,
  left=0mm,
  right=0pt,
}

\newcommand{\Set}[1]{\ensuremath{\{#1\}}\xspace}

\newcommand{\PPT}{\ensuremath{\mathrm{PPT}}\xspace}

\newcommand{\cind}{\ensuremath{\stackrel{\text{c}}{\approx}}\xspace}

\newcommand{\negl}{\ensuremath{\mathsf{negl}}\xspace}
\newcommand{\poly}{\ensuremath{\mathsf{poly}}\xspace}

\newcommand{\Adv}{\ensuremath{\mathcal{A}}\xspace}
\newcommand{\Challenger}{\ensuremath{\mathcal{C}}\xspace}
\newcommand{\SecPar}{\ensuremath{\lambda}\xspace}
\newcommand{\pick}{\ensuremath{\xleftarrow{\$}}}

\newcommand{\etal}{{et al.\ }}

\newcommand{\sysname}{\ensuremath{\mathrm{PDS}}\xspace}

\newcommand{\PD}{\ensuremath{\mathsf{PD}}\xspace}

\newcommand{\MultPD}{\ensuremath{\mathsf{MUL}\text{-}\mathsf{tPD}}\xspace}

\newcommand{\tPD}{\ensuremath{\mathsf{tPD}}\xspace}

\newcommand{\Setup}{\ensuremath{\mathsf{Setup}}\xspace}
\newcommand{\Oper}{\ensuremath{\mathsf{Oper}}\xspace}

\newcommand{\init}{\ensuremath{\mathsf{init}}\xspace}

\newcommand{\Dev}{\ensuremath{\mathcal{D}}\xspace}

\newcommand{\Key}{\ensuremath{\mathsf{K}}\xspace}

\newcommand{\pub}{\ensuremath{\mathsf{pub}}\xspace}
\newcommand{\hid}{\ensuremath{\mathsf{hid}}\xspace}
\newcommand{\Pub}{\ensuremath{\mathsf{Pub}}\xspace}
\newcommand{\Hid}{\ensuremath{\mathsf{Hid}}\xspace}

\newcommand{\Pat}{\ensuremath{\mathcal{P}}\xspace}
\newcommand{\Tra}{\ensuremath{\mathcal{T}}\xspace}

\newcommand{\state}{\ensuremath{\mathsf{st}}\xspace}

\newcommand{\Lay}{\ensuremath{\mathcal{L}}\xspace}
\newcommand{\FS}{\ensuremath{\mathsf{FS}}\xspace}
\newcommand{\BD}{\ensuremath{\mathsf{BD}}\xspace}
\newcommand{\FTL}{\ensuremath{\mathsf{FTL}}\xspace}
\newcommand{\FT}{\ensuremath{\mathsf{FT}}\xspace}

\newcommand{\LtPD}{\ensuremath{\mathcal{L}\textsf{-}\tPD}\xspace}
\newcommand{\LPD}{\ensuremath{\mathcal{L}\textsf{-}\PD}\xspace}

\newcommand{\Rule}{\ensuremath{\mathcal{R}}\xspace}

\newcommand{\wORAM}{\ensuremath{\mathsf{wORAM}}\xspace}

\newcommand{\Prot}{\ensuremath{\mathrm{\Pi}}\xspace}

\newcommand{\Read}{\ensuremath{\texttt{Read}}\xspace}
\newcommand{\Write}{\ensuremath{\texttt{Write}}\xspace}
\newcommand{\Erase}{\ensuremath{\texttt{Erase}}\xspace}

\newcommand{\Enc}{\ensuremath{\mathsf{E}}\xspace}
\newcommand{\bits}{\ensuremath{\{0,1\}}\xspace}

\newcommand{\Mount}{\ensuremath{\texttt{Mount}}\xspace}
\newcommand{\Unmount}{\ensuremath{\texttt{Unmount}}\xspace}

%% file: introduction.tex

\section{Introduction}

Data privacy has become essential maybe more so than at any other time in human history. Encryption can be used to defend against unauthorized disclosure of sensitive data, yet is not enough to handle adversaries empowered by law or rubber-hose (e.g.~oppressive governments) to coerce the user into revealing encryption keys. 

Unfortunately, numerous real-life examples show that protecting sensitive data in the presence of such coercive adversaries is often a matter of life and death. The Human Rights Group Network for Human Rights Documentation at Burma (ND-Burma)~\cite{DBLP:conf/ndss/PetersGP15} documented large numbers of human rights violations. Proof was carried out of the country on mobile devices by ND-Burma activists, risking exposure at checkpoints and border crossings. In 2012, a videographer could smuggle evidence of human rights violations out of Syria by hiding a micro-SD card in a wound on his arm~\cite{syria} etc. Threats of coercive attacks are not merely an Orwellian fantasy, but a real concern \cite{eg1,anderson1998steganographic,kennedy2000encryption,adv1,adv2,adv3,star2012syrian}.

To address this, {\em plausible  deniability}  (PD) has been proposed. It is a powerful  property,  enabling users to hide the existence of sensitive information on a system under inspection by overreaching or coercive adversaries, democratically  elected  or  otherwise. 

In the context of secure storage\footnote{PD has been first formalized in a (mostly theoretical) context of encryption \cite{canetti1997deniable,o2011bi}, often involving small amounts of data and sometimes read-only. This work focuses on applied aspects as they relate to efficient, modern, high-capacity data storage.}, PD refers to the ability of a user to plausibly deny the existence of certain data stored on a storage device even when an adversary has access to the device. Since adversaries cannot conclude anything about the existence of sensitive data, they have no good excuse to perform the coercion further, thus leaving those data in safety.

PD was first proposed in 1998 \cite{anderson1998steganographic}. Since then, popular encrypted file systems ({\FS}es) such as TrueCrypt \cite{Truecrypt} (first released in 2004) and other PD research results have emerged \cite{mcdonald1999stegfs,pang2003stegfs,skillen2013implementing,blass2014toward,DBLP:conf/ndss/PetersGP15,jia2017deftl} attempting to balance the ever present security-efficiency trade-off.



Unfortunately, existing efforts were designed for very specific adversaries and contexts, and under sometimes unclear security models and device assumptions. However, to achieve strong PD guarantees, it is important to understand and evaluate these contexts and limitations properly. 
%
%
This work aims to systematize knowledge and provide a more in-depth understanding for today's practitioners, and future research.

\subsection{Challenges}
\label{section:intro:challenges}
Before diving in, it is important to understand some of high-level challenges facing plausibly deniable systems researchers and practitioners.

\para{Security-Efficiency Trade-Off. Real-Life Adversaries.} 
Previous PD literature has been focusing on {\em single-snapshot} adversaries who can check the storage device only once, and {\em multi-snapshot} adversaries who can checks the device at several different time points. While the former are relatively easy to handle (proof being practical systems such as TrueCrypt \cite{Truecrypt}), {\em practical} PD systems resilient against multi-snapshot adversaries turns out to be more difficult to design. 

Ideally, researchers would like to obtain multi-snapshot security against all probabilistic polynomial time (PPT) adversaries\footnote{Security against all PPT adversaries is the golden rule for most cryptographic primitives and security tasks, e.g.~one-way functions, encryption schemes, digital signatures etc.} (referred to as ``full security''). However, until today, only a few constructions \cite{blass2014toward,chakraborti2017datalair,chen2019} achieve this level of security, but are unfortunately significantly slower than the underlying storage device. Other solutions seek better performance by relaxing the security requirements. For example, some of them assume that a small area on the device is hidden from the adversary, and some put certain restrictions on the adversarial behavior (see \Cref{section:comparison} for details).

Overall, unfortunately, no {\em practically efficient} construction achieves multi-snapshot PD with {\em full security}. 
This may be also because existing adversarial models and associated solutions have been developed mostly ad-hoc and not designed to answer more general, fundamental questions regarding the security-efficiency trade-off. For example, is there a performance bottleneck inherent to the concept of PD? Are $\wORAM$s necessary to achieve fully-secure PD? Are there multiple dimensions along which the PD security-efficiency trade-off can be optimized? We believe that answers to these questions are critical for both practitioners of today aiming to build in plausible deniability into modern system stacks, as well as for upcoming research in PD.

\para{Dependency on System Layers.}
To complicate things further, modern systems feature layered structures all of which persist state and can compromise any security guarantees aimed for by other layers. Consider that ubiquitous stack of a typical \FS, \FS caches, LVM layers, LVM caches, block-devices (\BD), block device caches, and flash translation layers (\FTL) (see \Cref{section:model:system} for additional details). Existing PD works consider only a specific layer, e.g.,  DEFY \cite{DBLP:conf/ndss/PetersGP15} builds PD in the \FS layer,  TrueCrypt \cite{Truecrypt} works in the \BD layer, DEFTL \cite{jia2017deftl} works in the \FTL layer. 

Further, most schemes make ad-hoc case-specific assumptions about the devices and the adversary behavior, accordingly achieving PD in a restricted sense. 

Such a layered structure complicates the security analysis. Schemes designed for a specific layer may lose their security guarantees if deployed at a ``wrong'' layer. As will be shown in \Cref{section:depedence-on-layers}, this fact can sometimes be overlooked unintentionally. Further, the existence of state in the other layers cannot be ignored since it often contains compromising information breaking the security of the overall scheme. 

In most cases, {\bf a realistic adversary with visibility into the state of one or more additional layers, may immediately compromise single-layer designs since the additional state can reveal access patterns and other security-sensitive information that a single-layer model simply cannot capture.}

It is thus critical to investigate the interplay between PD security and layers, and provide constructions and definitions with reduced or zero dependency on layers. Ideally, such an investigation can isolate PD as an independent security concept, and not only a layer/device-dependent property (\Cref{section:depedence-on-layers,sec:unified-def-mainbody}).

\para{Lack of Unified Security Framework.}
As discussed, full security as defined in \cite{blass2014toward} is achieved by only a few constructions which feature prohibitive performance overheads. Most other schemes restrict adversaries significantly and do not provide strong security or allow even for a comparative analysis of security. Very often also, the security arguments for such schemes contain heuristics, a very dangerous practice. For example, the security of DEFY \cite{DBLP:conf/ndss/PetersGP15} relied on the authors' claim that the hidden pages in their scheme were indistinguishable from secure-deleted public pages. However, with no formal proof given, it was not clear whether the asserted indistinguishability really held against all PPT coercive adversaries. Subsequently Jia et al.~\cite{jia2017deftl} showed that DEFY can be easily compromised with very little effort (if adversaries make several attempts to exhaust writing capacity).

Moreover, due to the lack of a unified security framework, different papers customize the definition of PD to serve their specific application or devices, making comparisons between systems difficult or outright impossible.   
This further leads to an unnecessary proliferation of threat models and definitions, with a polymorphous-yet-confusing naming style. For example, PD schemes deployed in the \FS layer are called ``steganographic file system'' or ``deniable file system'', while schemes designed for the \BD layer are named ``hidden volume encryption'' or ``deniable encryption''. In selecting a proper plausible deniability mechanism for their application, practitioners end up bewildered by such multifarious names, and the lack of structure or relationships among the security guarantees provided by those schemes.
It is essential to unify these adversarial definitions and application scenarios, and thus enable comparison-based evaluations.

\subsection{Contributions}
This work synthesizes existing ideas into a guide for system and security practitioners helping to understand, design or implement plausible deniability into new or existing systems. Concretely: 

\begin{enumerate}
\item
We observe that a key point of PD lies in concealing users' hidden data access patterns. Often this happens using randomized (ORAMs) or canonical form I/O. We examine how these approaches affect the security and efficiency of the resulting PD schemes. We also survey another approach appeared recently---basing the secrecy of access patterns on inherent properties of storage systems/devices. This approach usually leads to lightweight solutions that are ``native'' to the underlying systems/devices.


\item
We investigate the interplay between security assurances, adversarial models and modern multi-layer storage stacks. This reveals a set of general principles and definitions that can be deployed for better security-efficiency trade-offs.

\item
We propose the concept of trace-oriented security 
to enable the design and evaluation of schemes providing layer-independent security guarantees. We show that trace-oriented security was achieved (though not claimed explicitly) by a few existing constructions \cite{blass2014toward,chakraborti2017datalair,chen2019}. We show that this stronger security notion comes with a price---equivalence to write-only ORAMs. 

\item
We provide a way to unify and evaluate solutions under a single framework, where the main differences are expressed as constraints on the power of the adversary. Saliently, this unified point of view provides a framework for the comparison and evaluation of PD solutions. We present a taxonomy of security for existing constructions.
\item
Finally, we identify important under-explored areas, and suggest new directions for future research.
\end{enumerate}


%% file: appendices.tex
\vspace{-5pt}

\section{Unified PD Definition Equivalence}
\label{appendix:game:equivalence}

We now show that the unified PD definition in \Cref{section:definition} is equivalent to the one in~\cite{chakraborti2017datalair}, which allows {\em both} $\Pat_0$ and $\Pat_1$ to contain hidden requests.

First, it is easy to see that the definition in \cite{chakraborti2017datalair} is no weaker than the one defined in \Cref{section:definition}, because allowing hidden requests in {\em both} $\Pat_0$ and $\Pat_1$ only grants the adversary more power in the CAP game. So, the only thing left is to show that the definition in \Cref{section:definition} is no weaker than that in in \cite{chakraborti2017datalair}. Roughly speaking, this is true for the following reason. Consider a pair of \cite{chakraborti2017datalair}-type challenge $\Pat_1 \coloneqq \Pat_\pub^1 \cup \Pat_\hid$ and $\Pat'_1 \coloneqq \Pat_\pub^1 \cup \Pat'_\hid$, both of which contain hidden requests (but share the same public part). The security guaranteed by \Cref{figure:MulPD} says that a $\Pat_0 \coloneqq \Pat_\pub^1 \cup \Pat_\pub^2$ should be indistinguishable with $\Pat_1$, and also with $\Pat'_1$. Thus, it must the case that $\Pat_1$ and $\Pat'_1$ are indistinguishable. {In the following, we formalize the above intuition.}

To prove it formally, we need to show that if a PPT adversary $\Adv$ can win the CPA game defined in~\cite{chakraborti2017datalair} with non-negligible probability, then it can be efficiently converted into another PPT $\Adv'$ that wins the CPA game defined in \Cref{figure:MulPD} with non-negligible probability. We construct $\Adv'$ as follows. $\Adv'$ begins by picking a random bit $b' \pick \bits$ and then runs $\Adv$ internally. In the $i$-th ($i\in \{1, \ldots, r\}$) round, $\Adv$ will send a pair of challenge requests $\Pat_0$ and $\Pat_1$ (we emphasize that both $\Pat_0$ and $\Pat_1$ contain hidden requests). When this happens, $\Adv'$ sets $\Pat'_1 \coloneqq \Pat_{b'}$; and $\Adv'$ sets $\Pat'_0$ to the public part of $\Pat_0$ (or equivalently, the public part of $\Pat_1$). $\Adv'$ uses $\Pat'_0$ and $\Pat'_1$ as its $i^{th}$-round challenge requests for its own CPA game (i.e., the game defined \Cref{figure:MulPD}), and forwards the response from its challenger to the internal $\Adv$. At the end, if $\Adv$ guesses $\Adv'$'s $b'$ correctly, $\Adv'$ will output $1$; otherwise, $\Adv'$ outputs $0$.

It is easy to see that if the $b$ picked by $\Adv'$'s challenger (in the game specified in \Cref{figure:MulPD}) equals $1$, then the view of the internal $\Adv$ is identical to the case when it is participating in the CPA game in \cite{chakraborti2017datalair}. Since $\Pr[b = 1] = 1/2$, it follows that with probability $1/2$, the internal $\Adv$ will ``think'' that it is participating in the CPA game from \cite{chakraborti2017datalair}. Recall that we assume that $\Adv$ wins the \cite{chakraborti2017datalair} CPA game with some non-negligible probability $p$. Therefore, $\Adv'$ will win its own \Cref{figure:MulPD} game with probability $p/2$, which is also non-negligible.

\section{Write-Only ORAMs from Trace-Oriented PDs}
\label{sec:wORAM_from_PDs}

\vspace{-5pt}

\subsection{Write-Only ORAMs}

\vspace{-5pt}

\para{Notations.} A {\em data request} is a tuple $(\mathsf{op}, \mathsf{addr}, \mathsf{d})$, where $\mathsf{op} \in \Set{\mathtt{Read}, \mathtt{Write}}$ denotes a $\mathtt{Read}(\mathsf{addr})$ or a $\mathtt{Write}(\mathsf{addr}, \mathsf{d})$ operation, $\mathsf{addr}$ denotes the identifier of the block being read or written, and $\mathsf{d}$ denotes the data being written. For an ORAM scheme $\Prot$ and a sequence $\vec{y}=\Set{r_1, \ldots, r_n}$ of data requests, let $\mathsf{PhysicalAcc}^\Prot(\vec{y})$ denote the the physical access pattern that is produced by executing $\Prot$ on $\vec{y}$.

\begin{definition}[Write-Only ORAMs \cite{blass2014toward,roche2017deterministic,li2017write}]\label{def:wORAM}
An ORAM scheme is {\em write-only oblivious} if for any two sequences of data requests $\vec{y}_0$ and $\vec{y}_1$ containing the same number of $\mathtt{Write}$ requests, it holds that
$$\mathsf{WOnly}(\mathsf{PhysicalAcc}^\Prot(\vec{y}_0)) \cind \mathsf{WOnly}(\mathsf{PhysicalAcc}^\Prot(\vec{y}_1)),$$
where $\mathsf{WOnly}(\cdot)$ filters out the read physical accesses, and $\cind$ denotes {\em computational indistinguishability}.
\end{definition}

\begin{remark}\label{remark:wORAM_length}
In \Cref{def:wORAM} $\vec{y}_0$ and $\vec{y}_0$ may have different length\footnote{This is in contrast to standard ORAMs, which considers $\vec{y}_0$ and $\vec{y}_1$ of equal length, and requires the indistinguishabilty between the execution results without applying $\mathsf{WOnly}(\cdot)$.}; they are only required to contain the same number of $\mathtt{Write}$ requests. This stipulates that the execution of $\mathtt{Read}$ requests does not incur any physical writes: otherwise two sequences with different number of $\mathtt{Read}$ requests might be easily distinguished by checking the number of resulted physical writes.

\end{remark}

\subsection{Write-Only ORAMs from Trace-Oriented PD}

\para{The High-Level Idea.} 
In the security game of trace-oriented PDs, it is guaranteed that the writing traces resulted from two adversarially chosen access patterns $\Pat^1_\pub \cup \Pat_\hid$ and $\Pat^1_\pub \cup \Pat^2_\pub$
are computationally indistinguishable. 
In particular, this implies the existence of two ``universal'' public patterns $\Pat^1_{\pub}$ and $\Pat^2_{\pub}$ with the following property: for any hidden patterns $\Pat_{\hid}$, the \Write traces resulted from $\Pat^1_{\pub} \cup \Pat_{\hid}$ are indistinguishable with that from $\Pat^1_{\pub} \cup \Pat^2_{\pub}$
Given a PD scheme under the above restriction, a \wORAM can be implemented as follows: to perform a target operation $\alpha = (\mathsf{op}, \mathsf{addr}, \mathsf{d})$, it first loads $\alpha$ into $\Pat_{\hid}$, and then executes the PDS access algorithm on $\Pat^1_{\pub} \cup\Pat_{\hid}$, where $\Pat^1_{\pub}$ is the aforementioned universal public pattern. Thanks to the security of the PDS, the \Write traces of the execution of $\Pat^1_{\pub} \cup\Pat_{\hid}$ are indistinguishable from those of the execution of $\Pat^1_{\pub} \cup \Pat^2_{\pub}$, whichever $\alpha$ is hidden inside $\Pat_{\hid}$. This provides the hiding of \Write operations as required by $\wORAM$s. This idea is formalized in \Cref{alg:wORAM_from_PDS}.

\begin{AlgorithmBox}[label={alg:wORAM_from_PDS},float=!t]{Write-Only ORAM from Trace-Oriented PDS}
\begin{algorithmic}[1]
        \Procedure{$\mathsf{ORAM.Setup}$}{$1^\SecPar$} \label{alg:wORAM_from_PDS:Setup}
            \State $\Key_\pub, \Key_\hid, \Tra_\mathsf{init} \gets \mathsf{PDS.Setup}(1^\SecPar)$
            \State Initialize the device/memory blocks by executing $\Tra_{\init}$
        \EndProcedure
        \item[] 
        \Procedure{$\mathsf{ORAM.Access}$}{$\mathsf{op}, \mathsf{addr}, \mathsf{d}$} 
            \State $\Pat_\hid \coloneqq (r_1, \ldots, r_n)\gets \mathsf{HiddenGen}(n,\mathsf{op}, \mathsf{addr}, \mathsf{d})$
            \If{$\mathsf{op} == \mathtt{Read}$}\Comment{if this is a \texttt{Read} request} \label{alg:wORAM_from_PDS:if_read}
                \State $\Pat \coloneqq $\Set{R_\mathsf{dummy}}$ \cup \Pat_\hid$ \label{alg:wORAM_from_PDS:if_read_padding}
            \Else \Comment{if this is a \texttt{Write} request}
                \State $\Pat \coloneqq \Pat^1_\pub \cup \Pat_\hid$
            \EndIf
            \State $\Tra \gets \mathsf{PDS.Oper}(\Key_\pub,\Key_\hid, \Pat)$
            \State {\bf return} $\Tra$
        \EndProcedure
        \item[] 
        \Procedure{$\mathsf{HiddenGen}$}{$n, \mathsf{op}, \mathsf{addr}, \mathsf{d}$} \label{alg:wORAM_from_PDS:HiddeGen}
            \State $r_1 = (\mathsf{op}, \mathsf{addr}, \mathsf{d})$
            \For{$i = 1$ to $n$}
                \State $r_i = R_\mathsf{dummy}$ \Comment{Pad the access pattern with dummy requests}
            \EndFor
            \State \textbf{return} $(r_1, \ldots, r_n)$
        \EndProcedure
    \end{algorithmic}
\end{AlgorithmBox}


\para{ORAM Setup.} The setup procedure (\Cref{alg:wORAM_from_PDS:Setup}) simply runs the $\mathsf{PDS.Setup}$ to get the keys for public and hidden PD requests, and a sequence of commands $\Tra_{\init}$ that is meant to initialize the PD scheme. Once the commands in $\Tra_{\init}$ are executed on the underlying device/memory blocks, the ORAM system is ready to work.

\para{ORAM Access.} On input a request $\alpha = (\mathsf{op}, \mathsf{addr}, \mathsf{d})$, the $\mathsf{ORAM.Access}$ procedure first invokes a sub-procedure called $\mathsf{HiddenGen}$ (\Cref{alg:wORAM_from_PDS:HiddeGen}), which pads $\alpha$ with $n-1$ (same) dummy request $R_\mathsf{dummy}$. This ``padding'' is necessary for the following reasons. Recall that the construction wishes to execute $\alpha$ by loading it in the hidden part of some input pattern to $\mathsf{PDS.Oper}$. To leverage the security of $\mathsf{PDS}$, the hidden part must have length $n$. This is exactly the purpose of $\mathsf{HiddenGen}$. Now, the procedure can create a pattern $\Pat = \Pat^1_\pub \cup \Pat_{\hid}$ by concatenating the ``universal'' $\Pat^1_\pub$ with output $\Pat_{\hid}$ of $\mathsf{HiddenGen}$; the writing traces for $\Pat$ will be indistinguishable with that for $\Pat^1_\pub \cup \Pat^2_{\pub}$, due to the security of $\mathsf{PDS}$. 

As mentioned in \Cref{remark:wORAM_length}, write-only ORAMs inherently require that $\mathtt{Read}$ request should not lead to physical writings. However, this condition may not be satisfied by the underlying $\mathsf{PDS}$. To see that, consider a $\mathtt{Read}$ request $\alpha$. Following the above strategy, the procedure will load $\alpha$ into $\Pat_\hid$ and set $\Pat = \Pat^1_\pub \cup \Pat_\hid$. Since $\Pat_\hid$ contains only $\alpha$ and some dummy requests, we can assume that  $\Pat_\hid$ does not cause any physical writings. However, the $\Pat^1_\pub$ part may contain some requests which incur physical writes. To resolve this issue, replace $\Pat^1_\pub$ with (the sequence of) a single dummy operation (\Cref{alg:wORAM_from_PDS:if_read_padding}), if $\alpha$ is a $\mathtt{Read}$ request. Since a dummy operation does not cause any writes, $\alpha$ can be executed without incurring physical writes.





\vspace{-5pt}
\begin{lemma}\label{lemma:tPD_wORAM}
If $\mathsf{PDS} = (\mathsf{Setup}, \mathsf{Op})$ is a secure PD scheme, then \Cref{alg:wORAM_from_PDS} is a secure write-only ORAM.
\end{lemma}\vspace{-5pt}

\begin{proof}
Let $\vec{y}_0$ and $\vec{y}_1$ be two arbitrary data request sequences that contain the same number of $\mathtt{Write}$ operations. Note that it is possible that $|\vec{y}_0|\ne |\vec{y}_1|$. For $b\in \Set{0,1}$, let $\mathsf{Out}_b$ denote the sequence of traces resulted from executing \Cref{alg:wORAM_from_PDS} sequentially on each requests in $\vec{y}_b$. The following shows that $\mathsf{WOnly}(\mathsf{Out}_0)$ and $\mathsf{WOnly}(\mathsf{Out}_1)$ are computationally indistinguishable.

Let $m$ denote the number of write operations in $\vec{y}_0$ (or $\vec{y}_1$). 
Note that $\mathsf{Out}_0$ and $\mathsf{Out}_1$ may have different length, because the length of $\mathsf{Out}_b$ depends on $\vec{y}_b$. But it is clear that $|\mathsf{WOnly}(\mathsf{Out}_0)|=|\mathsf{WOnly}(\mathsf{Out}_1)| = m$ due to the following two facts:
\begin{enumerate}
    \item by construction (specifically, \Cref{alg:wORAM_from_PDS:if_read} and \Cref{alg:wORAM_from_PDS:if_read_padding}), $\mathtt{Read}$ requests do not cause any writing traces;
    \item both $\vec{y}_0$ and $\vec{y}_1$ contain exactly $m$ $\mathtt{Write}$ requests.
\end{enumerate}  
Moreover, by the security of $\mathsf{PDS}$, running \Cref{alg:wORAM_from_PDS} on any $\mathtt{Write}$ request has the same effect of executing $\mathsf{PDS.Oper}(\Key_\pub,\Key_\hid, \Pat^1_\pub \cup \Pat^2_\pub)$. Therefore, it follows that for any $b \in \Set{0,1}$, 
\begin{equation}\label{eq:trace_out_b}
\underbrace{\big(\mathsf{WOnly}(\Tra^*), \ldots, \mathsf{WOnly}(\Tra^*)\big)}_{\text{repeat $m$ times}} ~~\cind~~ \mathsf{WOnly}(\mathsf{Out}_b),
\end{equation}
where $\Tra^*$ denotes the output of the following operation:
$$\mathsf{PDS.Oper}(\Key_\pub,\Key_\hid, \Pat^1_\pub \cup \Pat^2_\pub).
$$ 
It then follows immediately from \Cref{eq:trace_out_b} that
$
 \mathsf{WOnly}(\mathsf{Out}_0)  \cind  \mathsf{WOnly}(\mathsf{Out}_1)
$. 
\end{proof}